\documentstyle[aps,twocolumn,floats,epsfig]{revtex}  

\draft
\begin{document}

\twocolumn[\hsize\textwidth\columnwidth\hsize\csname
@twocolumnfalse\endcsname

\title{High pressure phases in highly piezoelectric PbZr$_{0.52}$Ti$_{0.48}$O$_{3}$}

 \author{A. Sani$^{1}$, B. Noheda$^{2}$, I. A. Kornev $^{3}$, L. Bellaiche$^{3}$, P. Bouvier$^{4}$, and J. Kreisel$^{5}$}
\address{$^{1}$European Synchrotron Radiation Facility, BP 220, F-38043 Grenoble France}
\address{$^{2}$Condensed Matter Physics, Vrije Universiteit, De Boelelaan 1081, Amsterdam 1081HV, The Netherlands.}
\address{$^{3}$Physics Department, University of Arkansas, Fayetteville, AR 72701, Arkansas, US}
\address{$^{4}$Lab. d'Electrochimie et Physicochimie des Mat\'{e}riaux, ENSEEG, BP 75, 38402 St. Martin d'H\`{e}res, France}
\address{$^{5}$Lab. des Mat\'{e}riaux et du G\'{e}nie Physique, ENS de Physique de Grenoble, BP 46, 38402 St. Martin d'H\`{e}res, France}

\date{\today}

\maketitle

\begin{abstract}
Two novel room-temperature phase transitions are observed, via synchrotron x-ray diffraction and Raman spectroscopy, in the
PbZr$_{0.52}$Ti$_{0.48}$O$_{3}$  alloy under hydrostatic pressures up to 16 GPa. A monoclinic (M)-to-rhombohedral (R$_{1}$) phase
transition takes place around 2-3 GPa, while this R$_{1}$ phase transforms into another rhombohedral phase, R$_{2}$, at $\simeq$ 6-7 GPa.
First-principles calculations assign the R3m and R3c symmetry
to R$_{1}$ and R$_{2}$, respectively, and reveal that R$_{2}$ acts as a pressure-induced structural bridge between the polar
R3m and a predicted antiferrodistortive R$\overline{3}$c phase.
 \end{abstract}

\pacs{PACS:64.70.Kb,77.80.Bh,62.50.+p}

\vskip2pc]

\narrowtext

\marginparwidth 2.7in
\marginparsep 0.5in

PbZr$_{1-x}$Ti$_{x}$O$_{3}$ (also called PZT) solid solutions with x $\simeq $ 0.5 are used, already
for decades, in virtually all piezoelectric devices from ultrasound generators to micropositioners, due
to their outstanding electromechanical performance~\cite{Jaff}. All PZT compositions have the cubic perovskite
structure at high temperatures, while they undergo a phase transition, at about 650 K, to a ferroelectric
rhombohedral R3m phase (with a polarization
pointing along the [111] pseudo-cubic direction) for Zr-rich samples  and to a ferroelectric tetragonal P4mm phase
(with a polarization lying along [001]) for Ti-rich samples~\cite{Jaff,Shir}. Recently, higher resolution synchrotron x-ray experiments at low temperatures in high quality samples
have made possible the resolution of the region around the R3m-P4mm boundary, and have revealed the
existence of a third ferroelectric phase with a lower monoclinic (M) Cm symmetry within a narrow
compositional range~\cite{Nohe1}. This M phase can be considered as a structural bridge between the R3m and P4mm phases since its polarization continuously rotates in the ($\bar{1}$10) plane
from the [111] to the [001] pseudo-cubic directions, as the Ti content increases~\cite{Guo}. Raman
measurements in PZT have also found mode splitting which are consistent with monoclinic
symmetry~\cite{Fran,Lima}, and first-principles calculations have shown that the polarization rotation
is responsible for the large piezoelectric coefficients observed in PZT~\cite{Bell,Cohe,Wu}.

Furthermore, antiphase oxygen rotations, superimposed on polar displacements, have been recently observed within the monoclinic phase of PZT at low temperature~\cite{Nohe3,MLT}. This is consistent with first-principles calculations predicting that oxygen rotations and polar displacements are close in energy~\cite{Forn}. Interestingly, the
energetical order between different phases in perovskite materials can be modified by applying an external pressure.
This is clearly evidenced by earlier works observing that pressure lowers the Curie temperature of phase transitions
induced by softening of zone-center phonons, like the $\Gamma _{15}$ polar modes in PbTiO$_{3}$~\cite{Sam1} --- and thus leads to a cubic
paraelectric phase at high enough pressure, as recently reported for PbZr$_{0.52}$Ti$_{0.48}$O$_{3}$~\cite{Rouq} ---
while it increases the temperature of the transitions related to zone-boundary modes like the antiferroelectric mode in PbZrO$_{3}$~\cite{Sam2}.
These opposite effects ~\cite{Forn,Sam1}, the fact that
Cm is a structure bridging the R3m and P4mm phases (and is thus very sensitive to a small change of interactions~\cite{Aaron}), and
the observed coexistence of polar displacements and oxygen rotation suggest that the phase diagram of PZT versus pressure
can be extraordinary rich.

Motivated to precisely determine {\it and} fully understand this phase diagram, we combined three complementary techniques, synchrotron x-ray diffraction, Raman spectroscopy and first principles calculations, to
investigate PbZr$_{0.52}$Ti$_{0.48}$O$_{3}$ under hydrostatic pressure. These studies revealed (1) two novel phase transitions, (2) two different pressure-induced bridging structures, and (3) the absence of a high pressure cubic paraelectric phase.

The same PbZr$_{0.52}$Ti$_{0.48}$O$_{3}$ ceramic pellets of Ref.~\cite{Nohe1} were used. A
piece was crushed into powder and loaded in a diamond anvil cell using both nitrogen and a
4:1 methanol-ethanol-mixture as pressure media for the x-ray and Raman measurements, respectively. A maximum
pressure of 16 GPa was reached in hydrostatic conditions. The pressure was measured by the ruby fluorescence method~\cite{Mao}. X-ray diffraction data were taken at room temperature in the ID9 beamline at the ESRF with a monochromatic beam of $\lambda$ =0.414 \AA\ focused to a 30x30 $\mu m^{2}$ spot and  using an angle dispersive set-up with image plate detector. The unit cells for the different pressures were obtained by means of a  full profile Le Bail analysis~\cite{LeBa,Riet}, using a pseudo-voigt peak shape with asymmetry correction~\cite{shap}. A second
 cubic phase was used to account for the diffuse scattering found in the monoclinic phase, as in ref.~\cite{Nohe1}. Depolarized Raman
 spectra of the sample powders were recorded at room temperature in back-scattering geometry with a Jobin Yvon T64000 spectrometer
 equipped with a microscope objective and the 514.5 nm line of an Ar+ ion laser as excitation line. The Raman spectra after pressure
 release were identical to the initial spectra attesting the reversibility of pressure-induced changes.

\begin{figure}
\epsfig{width=0.7\linewidth,figure=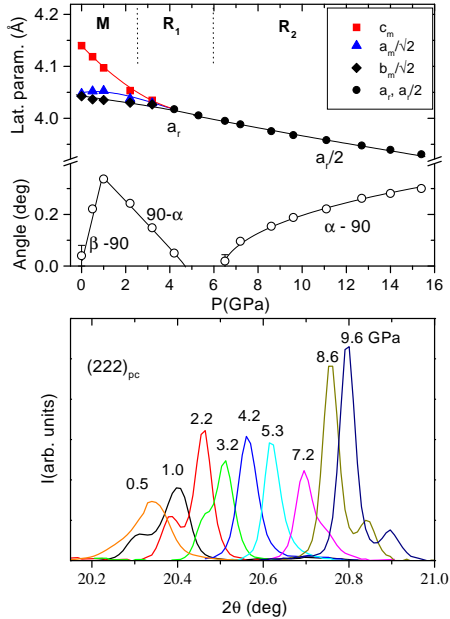}
\caption{(Color online) Evolution of the lattice parameters with increasing pressure (Top).
Solid lines are a guide to the eye. Evolution of the (222) x-ray
diffraction peak with pressure (Bottom).}
\end{figure}

As shown in Figure 1(top), at atmospheric pressure PbZr$_{0.520}$Ti$_{0.48}$O$_{3}$ is just at
the monoclinic-tetragonal phase transition
showing a very small monoclinic distortion, with lattice constants
a$_{m}$, b$_{m}$, c$_{m}$ and a $\beta $ angle. Pressure induces fairly large and continuous changes in
$c_{m}$, in agreement with ref.\onlinecite{Rouq}, and far more subtle changes in a$_{m}$, b$_{m}$ and $\beta $.
 Small pressures below 2 GPa increase the monoclinic
distortion by increasing both $\beta$ and the difference between a$_{m}$ and b$_{m}$. At $\sim $ 2-3 GPa,
a$_{m}$= b$_{m}$=c$_{m}$, and the triplet of the pseudo-cubic (222) reflection (see Fig. 1 (bottom) ) observed at low pressures changes into a doublet. The corresponding phase, denoted by R$_{1}$, is rhombohedral with an angle $\alpha $ $<$ 90$^{o}$ , or unit cell elongated along [111]~\cite{doub}. The most plausible space group for R$_{1}$ is R3m since this latter is ferroelectric with polarization along [111]. For P$>$ 3 GPa, the rhombohedral distortion of R$_{1}$  decreases with increasing pressure, until at about 5 GPa when the diffraction pattern seems
to be cubic with no visible peak split in agreement with ref.\onlinecite{Rouq}.
However, an analysis of the peak widths shows that they are not resolution limited, suggesting local non-cubic distortions.
Interestingly, when the pressure is further increased, above 7 GPa,
 the distortion from cubic continuously increases and the diffraction peaks start splitting again. As seen  in Figure 1(bottom), the diffraction pattern is again that of a rhombohedral phase but this time
 with $\alpha $$>$ 90$^{o}$~\cite{doub}. This new phase (denoted by 
R$_{2}$) is thus associated to a unit cell flattened along [111]. One good candidate for the symmetry of R$_{2}$ is R$\overline{3}$c since all known rhombohedral perovskites with $\alpha $$>$ 90$^{o}$ adopt this space group\cite{Mega}, a well-known example being LaAlO$_{3}$. In this {\it non-polar} structure the distortion from cubic is due to anti-phase rotations of the oxygen octahedra around the [111] direction, which results from the condensation of zone-boundary modes (R$_{25}$) and thus double the unit cell~\cite{Axe}. This type of rotations is also observed in Zr-rich PZT at atmospheric pressure and low temperature, although in this case polar displacements are also present and the structure has the space group R3c with $\alpha $$<$90$^{o}$~\cite{Glaz}.

\begin{figure}
\epsfig{width=0.7\linewidth,figure=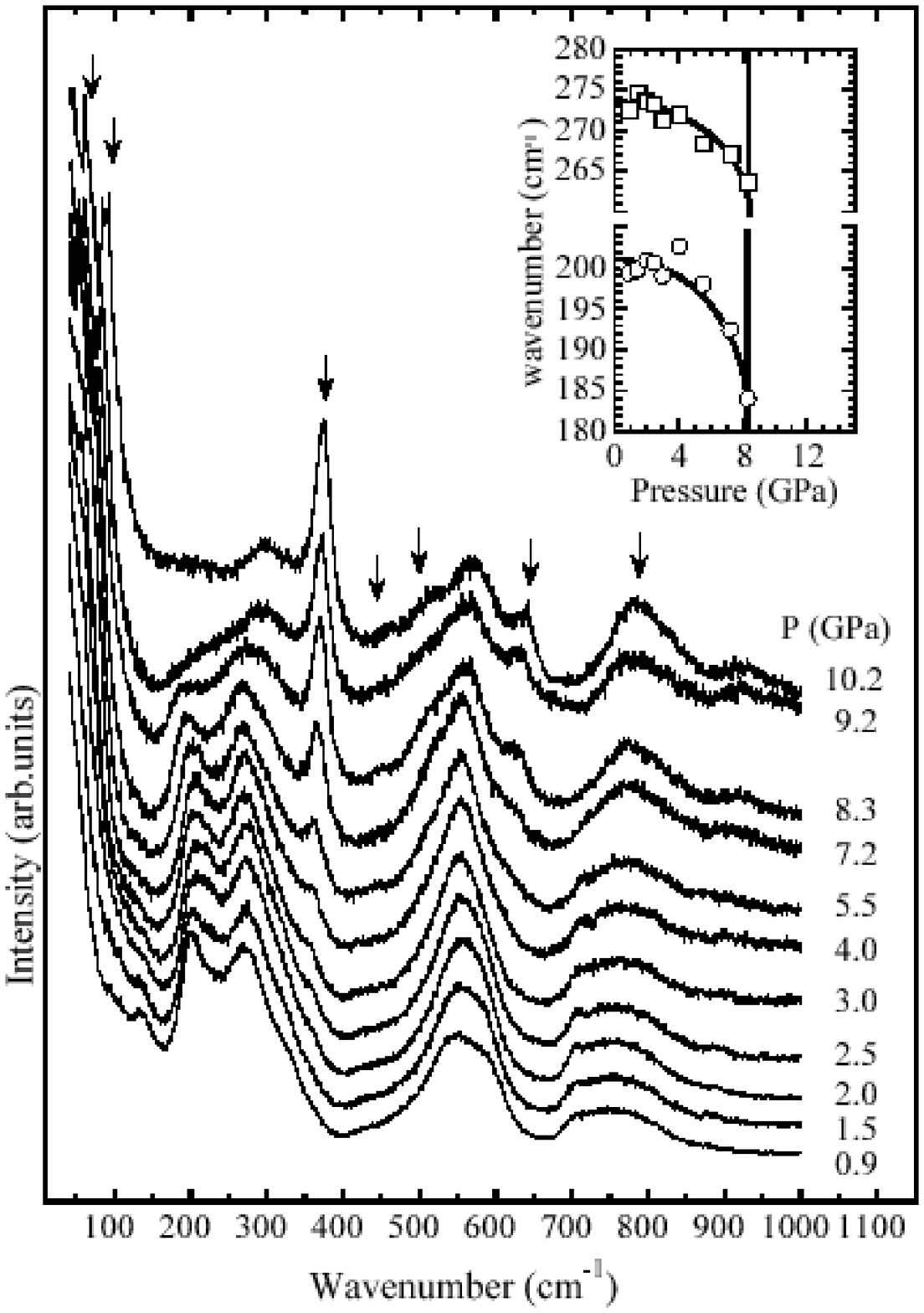}
\caption{Evolution of the Raman spectra with increasing pressure. Arrows
mark new reflections in the high pressure phase. The inset shows the softening of
the modes at 200 and 280 cm$^{-1}$ at the R$_{1}$-R$_{2}$ phase transition.}
\end{figure}

Raman measurements were performed to complement the x-ray data.
Figure 2 presents the evolution of the Raman spectra of the same sample with increasing pressure
until 10.2 GPa. A particularly striking finding is that the Raman signal is fairly well-defined for any pressure, including the ones around 5-7 GPa for which x-ray shows a cubic-like structure, implying the existence of non-cubic distortions on the Raman characteristic length scale, too small for x-rays to resolve it. Furthermore, the
spectra at low pressures are similar to those reported in the recent literature~\cite{Fran,Lima,Rouq}
for similar compositions.  Several changes are noticed upon increasing pressure. The most obvious one being the appearance of several new bands at high pressures (marked with arrows in the figure) and the suppression of some others. Most of the changes appear in between 5.5 and 7.2 GPa, in agreement with the critical pressure observed by x-ray diffraction
for the transition to the R$_{2}$ phase. The strong and sharp band at 380 cm$^{-1}$ has been observed under
pressure in other perovskites and has been associated with rhombohedral symmetry~\cite{Krei1,Krei2}.
Its narrow width at high pressures is indicative of the long-range character of the distortion, and the
increase in the number of bands is also consistent with a doubling of the unit cell. Furthermore, we observe a softening (inset Figure 2) and progressive disappearing of the two modes in the 200-to-350 cm$^{-1}$ range, clearly supporting the occurence of a phase transition, most possibly related to fundamental changes of the Zr/Ti-cation displacement.

\begin{figure}
\epsfig{width=0.6\linewidth,figure=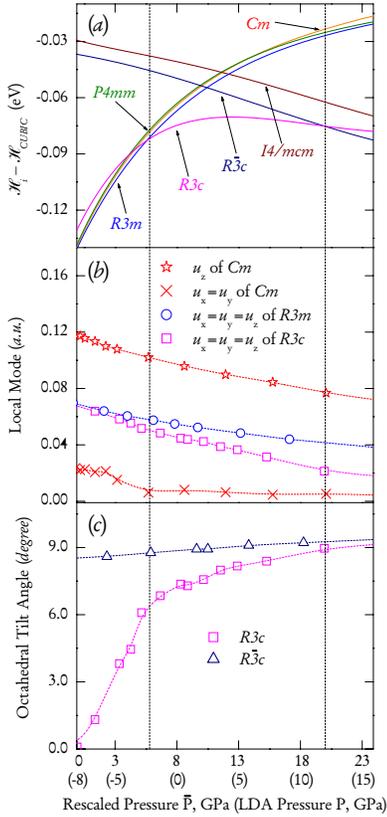}
\caption{(Color online) First-principles prediction of the pressure behavior of (a) $\Delta H$ (see text) for all
the considered phases (the phase corresponding to a minimal  $\Delta H$ at a given $P$ is thus the
most stable one for this pressure), (b) the polar local soft-mode for the Cm, R3m and R3c phases, and
(c) the rotation angle of the oxygen octahedra with respect to the pseudo-cubic [111] direction for
the R3c and R$\overline{3}$c phases. $\bar{P}$ is the rescaled pressure (see text). LDA-pressure P is indicated in parenthesis}
\end{figure}

In order to gain further insights, we also perform 0K calculations within the local density-approximation (LDA)~\cite{LDA}, using the virtual crystal approximation (VCA)
proposed in Ref.~\cite{VCA} (in which further details of the calculations can be found). We limit
ourselves to the study of 7 phases: the {\it paraelectric} cubic Pm-3m state; the {\it ferroelectric}
monoclinic Cm, rhombohedral R3m and tetragonal P4mm phases; the {\it antiferrodistortive}
rhombohedral R$\overline{3}$c   and tetragonal I4/mcm phases; and the rhombohedral R3c state that can exhibit {\it both} polar and antiferrodistortive degrees of freedom. The lattice vectors of the Cm, R3m and P4mm phases are those experimentally determined in Refs.~\cite{NohedaCmP4mm,NohedaR3m} for an atmospheric pressure, and the associated  strain tensors -- with respect to the cubic case -- are kept frozen in our calculations. On the other hand, for computational convenience, the lattice vectors of the R$\overline{3}$c, I4/mcm and R3c phases do not contain any strain with respect to the cubic phase.
Fig~3a shows  the $\Delta H=H-H_{cubic}$ difference between the enthalpy of each phase and the $H_{cubic}$ enthalpy of the cubic paraelectric phase as a function of pressure. It is well known that
LDA usually underestimates the lattice constant, which explains why the LDA-predicted pressure corresponding to the experimental volume determined in Ref~\cite{NohedaCmP4mm} for the Cm phase, at
low temperature and at an atmospherique pressure, is found to be $-$8 GPa. Therefore, in order to compare with the experiments, we have performed an uniform shift of the theoretical $P$ scale by 8
GPa, as indicated in Figs~3 by the {\it rescaled} theoretical pressure $\bar{P}$.

One can notice that our simulations yield a R3m phase that is slightly more favorable than Cm for $\bar{P}=0$. This discrepancy between theory and experiments for the ground-state symmetry is likely due to the fact that we neglect an important feature of monoclinic phases in our VCA calculations, namely the significant disorder existing between local dipoles located in different 5-atom cells~\cite{Rapp,Aaron1}. On the other hand, further using this rescaled pressure scale leads to a R3m--R3c transition occurring around 6 GPa. This is in remarkable agreement with the
experimentally-determined R$_{1}$--R$_{2}$ transition pressures mentioned above~\cite{footnoteP}, and thus strongly suggests that the  spaces groups of R$_{1}$ and R$_{2}$ are R3m and R3c, respectively. The
calculations further predict that R3c transforms into a R$\overline{3}$c  phase for pressure $\bar{P}$ around 20 GPa. (Note that  $\Delta H$ of the R3c and R$\overline{3}$c phases are quite negative even for $\bar{P}$ larger than 20 GPa, which further confirms the {\it absence }of a high pressure cubic phase.)

Fig~3b and Fig~3c show the pressure behavior of the local soft-mode ${\bf u}$ (directly related to the spontaneous polarization~\cite{Zhong}) and the rotation angle of the
oxygen octahedra, respectively, for the relevant phases. One can clearly see that the z-coordinate of ${\bf u}$ in the Cm phase significantly decreases when increasing $\bar{P}$ from 0 to 2
GPa, while the corresponding x- and y-components are barely affected by this change of pressure. (Note that the x-, y- and z- axes are chosen along the pseudo-cubic [100], [010] and [001] directions,
respectively). These simulated behaviors are consistent with the pressure changes of a$_m$, b$_{m}$ and c$_{m}$ depicted in Fig~1 for the M phase, since ferroelectrics are well-known to exhibit a
coupling between polarization (related to  ${\bf u}$) and strain (related to lattice
vectors)~\cite{Zhong}. Similarly, the continuous decrease predicted for every component of ${\bf u}$ in the R3m phase  -- when increasing  $\bar{P}$ from 2 to 6 GPa -- is consistent
with the experimentally observed increase (see, Fig~1)  towards  $90^{o}$ for the rhombohedral angle of R$_{1}$ in this range.
 Figs~3b and 3c further indicate that in the R3c phase for rescaled pressures above 3 GPa, the spontaneous polarization decreases in magnitude while the rotation of the oxygen octahedra increases, at increasing pressure. First-principles calculations thus provide a successful explanation for the unusual phase that is
Raman-active but characterized by a $\alpha$ angle $\simeq 90^{o} $ between its lattice vectors, and that is observed for pressure ranging between 5 and 7 GPa (see Figs~1 and 2): this phase is simply a R3c structure in which polar displacements (favoring $\alpha < 90^{o}$ ) compete with the rotation of
the oxygen octahedra (favoring $\alpha > 90^{o}$). When further increasing pressure, the simulations yield a significant increase (decrease) of the antiferrodistortive (polar) structural features. This explains why $\alpha$  increases with pressure after 7 GPa (see Fig.1), and also indicates that R3c behaves as a pressure-induced bridging structure between the R3m and R$\overline{3}$c phases (which are not in a simple super- or subgroup relation).

In summary, complementary synchrotron x-ray diffraction, Raman spectroscopy and first principles
calculations in piezoelectric PbZr$_{0.52}$Ti$_{0.48}$O$_{3}$, under pressure, reveal two novel room-temperature
phase transitions. The first transition occurs around 2-3 GPa from the ferroelectric
monoclinic Cm phase to a ferroelectric rhombohedral R3m phase. The second transition occurs at $\sim $6-7 GPa, from the ferroelectric rhombohedral R3m phase to a rhombohedral R3c phase exhibiting both
polar and antiferrodistortive structural features. The first transition thus involves polarization
rotation, while the second transition leads to a smooth suppression of the polar displacements and
continuously increases the oxygen rotations around the Zr/Ti cations with increasing pressure. The
pressure phase diagram of PZT thus contains two peculiar bridging structures: polar Cm that can be
seen as the natural evolution of polar P4mm to polar R3m, and R3c that brings polar R3m to
antiferrodistortive R$\overline{3}$c. The absence of a high-pressure cubic phase is, therefore, reported. Our results may also be relevant to thin films because of the crucial role played by the stress in these two-dimensional systems.
Finally, we would like to raise the possibility that the mysterious (cubic-like while polar) so-called X-phase, recently reported in some complex perovskites~\cite{Xu}, is a phase similar to the one we observe at the R3m-R3c transition, namely a phase in which polar distortions compete with antiphase oxygen rotations.

We would like to thank R. Guo and the late S.K. Eagle Park for the excellent samples and D.E. Cox and G. Shirane for very useful discussions.
The theoretical part of this work is supported by Office of Naval Research Grants
N00014-01-1-0600 and N00014-01-1-0365 (Center for Piezoelectrics by Design),
and National Science Foundation Grant DMR-9983678.


\begin{references}
\bibitem{doub} A doublet with intensity ratio 1:3 (3:1) is expected for the
(hhh) reflection of a Rhomboedral phase with $\alpha $$<$ 90$^{o}$ ($>$ 90$^{o}$).
\bibitem{footnoteP} The uncertainty of the theoretical transition pressure
between R3m and R3c is estimated to be $\pm$ 1 GPa.
\end{references}
\end{document}